\documentclass[12pt,draftcls, onecolumn,journal]{IEEEtran}
\usepackage{cite,graphicx,amssymb,amsmath,color,textcomp}
\usepackage{extarrows,multirow,multicol}
\usepackage{bm}
\usepackage{subeqnarray}
\usepackage{amsthm}
\usepackage{amsfonts}
\usepackage{multirow}
\usepackage{tabularx}
\usepackage{subfig}
\usepackage{verbatim}
\usepackage{stfloats}

\newtheorem{remark}{\underline{Remark}}%[section]

\newtheorem{theorem}{Theorem}
\newtheorem{corollary}[theorem]{Corollary}
\newtheorem{lemma}[theorem]{Lemma}

\newlength{\figwidth}
\begin{document}
\title{ 
{Performance Analysis of a Two--Tile Reconfigurable Intelligent Surface Assisted  $2\times 2$ MIMO System}
}
\author{Prathapasinghe~Dharmawansa, {\it  Member, IEEE}, \\Saman~Atapattu, {\it Senior Member, IEEE}, and  Marco Di Renzo, {\it  Fellow, IEEE}%, and Jamie~Evans, {\it Senior Member, IEEE}
\thanks{
 P.~Dharmawansa is with Department of Electronic and Telecommunications Engineering, University of Moratuwa, Moratuwa, Sri Lanka (email: prathapa@uom.lk).}
 \thanks{
S.~Atapattu is with the Department of Electrical and Electronic Engineering,
The University of Melbourne, Victoria, Australia, (e-mails: saman.atapattu@unimelb.edu.au). }
\thanks{
M. Di Renzo is with Universit\'e Paris-Saclay, CNRS and CentraleSup\'elec, Laboratoire des Signaux et Syst\`emes,  Gif-sur-Yvette, France. (e-mail: marco.direnzo@centralesupelec.fr)
}
\vspace{-10mm}
}
\maketitle
\begin{abstract}

We consider a two--tile reconfigurable intelligent surface (RIS) assisted wireless network with a two-antenna transmitter and receiver over Rayleigh fading. We show that the average received signal-to-noise-ratio (SNR) optimal transmission and combining vectors are given by the left and right singular spaces of the RIS-receiver and transmit-RIS channel matrices, respectively. Moreover, the optimal phases at the two tiles of the RIS are  determined by the phases of the elements of the latter spaces. To further study the effect of phase compensation, % at the RIS, 
we statistically characterize the average SNR of all possible combinations of transmission and combining directions pertaining to the latter singular spaces by deriving novel expressions for the outage probability and throughput of each of those modes. Furthermore, for comparison, we derive the corresponding expressions in the absence of RIS. Our results show an approximate SNR improvement of $2$\,dB due to the phase compensation at the RIS.
%A significant outage and throughput gains due to the RIS are also observed via our analytical results. 

%We consider a fading channel in which a two-antenna transmitter communicates with a two-antenna receiver
%through a two-element reconfigurable intelligent surface (RIS).
%The beamforming vector at the transmitter, the combining vector at the receiver, and the phase shifts of the RIS are optimized in order to maximize the average signal-to-noise-ratio (SNR) at the receiver. 
%By assuming Rayleigh fading, we prove that the SNR ....
%The proposed analytical framework allows us to quantify the robustness of RIS-aided transmission to fading channels. 
%For example, we prove that the ....

\end{abstract}


\newpage
\section{Introduction} \label{S1}

Reconfigurable intelligent surfaces (RISs) have been identified as one of the possible physical-layer technologies for beyond-5G networks. RIS can passively beamform the received signal  from a wireless transmitter towards its receiver by using large arrays of antenna elements which are usually spaced half of the wavelength apart~\cite{renzo2019corr}. 
%RIS is a low power and low cost technology which does not require power amplifiers, and thereby has a reduced  signal range as a drawback. 
Since the benefit of RISs can only be achieved by properly configured  phase shifts of passive reflective elements in real time, most existing work on RISs focus on phase optimization, see e.g.,~\cite{Wu2019twc,zappone2020overheadaware,renzo2020smart} and references therein. %,Huang2019twc,Guo2020twc,Huang2020jsac}. 
In contrast, communication-theoretic performance limits of RIS assisted single-antenna systems have been analyzed in \cite{Han2019tvt,Nadeem2020twc,Jung2020twc,%Basar2019acc,
Zhao2020wcoml,Atapattu2020tcom,%Boulogeorgos2020acc,
Zhang2020vt }.
%As far as the performance analysis of RIS-aided systems is concerned,  there are several works, that explored the communication-theoretic performance limits of single-antenna systems \cite{Han2019tvt,Nadeem2020twc,Jung2020twc,Basar2019acc,Zhao2020wcoml,Atapattu2020tcom,Boulogeorgos2020acc,Zhang2020vt }. 
However, few research efforts explored the performance of multi-antenna systems \cite{Di2020jsac}.
%because optimum phase angles are returned by optimization algorithms, which are usually numerical values. Therefore, analytical performance analysis is not feasible.  

%For single-antenna end users, in particular, performance analysis in terms of the ergodic capacity, spectral efficiency and error rate are investigated %by characterizing the receive signal-to-noise ratio (SNR) using the central limit theorem (CLT), e.g.,
%in~\cite{Han2019tvt,Nadeem2020twc,Jung2020twc,Basar2019acc,Zhao2020wcoml,Atapattu2020tcom,Kudathanthirige2020arxiv,Boulogeorgos2020acc}. %Recently, a more accurate performance analysis framework  is proposed for a general one-way and two-way RIS system in \cite{atapattu2020wcnc,Atapattu2020tcom}, where a single product-channel is characterized by a gamma random variable. Asymptotic analysis shows that the outage decreases at $\left(\log(\rho)/\rho\right)^L$ rate where $L$ is the number of elements, whereas the spectral efficiency  increases at $\log(\rho)$ rate at large average signal-to-interference-plus-noise  ratio SINR $\rho$. 
%Moreover, the impact of phase noise has been investigated in \cite{Badiu2020coml,qian2020beamforming}
%is quantified in \cite{Badiu2020coml}, and applied in \cite{Ferreira2020ojcomsoc,qian2020beamforming}. %Moreover, in \cite{Zhao2020coml,Zhao2020wcoml}, the authors integrate a backscatter link with a RIS to enhance backscatter communications, and  analyze the SER performance. 
For a large number of elements at the RIS, the authors of \cite{qian2020beamforming} investigate an RIS-assisted multiple-input and multiple-output (MIMO) network.
%where the transmitter and receiver are equipped with  multiple antennas.
%and RIS consist of multiple antennas or elements. 
%A common trend in large system analysis, which heavily depends on the concentration of probability measures, has its usefulness  when finite dimensional analysis is intractable. 
%Since we can readily apply sophisticated properties in probability theory, such as  the law of large numbers, to circumvent analytical difficulties, performance analysis of a large number of antennas/elements  systems is common in the literature.
%However, the  simplest two--by--two MIMO system is  of paramount importance  for   current and future wireless systems, because each of access points as well as users may have only one,  two or very few antennas. % \cite{Liu2012mag}. 
%Moreover, a two-element RIS can also be viewed as  two RISs (or two tiles) acting as anomalous or conventional reflectors whose phases can be adjusted independently \cite{najafi2020physicsbased}. 
Exact analytical characterization of  finite dimensional RIS-assisted systems, on the other hand, is not available in the current literature.
%Due to the analytical difficulties, this problem has not been investigated so far. 
Therefore, different from previous works, this paper analytically characterizes the exact performance, in terms of outage probability and throughput, of a fundamental finite dimensional RIS-assisted MIMO system. % where end-users and RIS consist of two antennas/elements. % over Rayleigh fading. 
In particular, we focus our attention on a $2\times 2$ MIMO system model 
%in which the transmitter and the receiver consist of two antenna elements, 
assisted by an RIS  made of two tiles that can be configured to operate as anomalous or conventional reflectors by adjusting independently their surface phase shifts \cite{najafi2020physicsbased}.  

We propose an analytical framework 
based on the representation of the $2\times 2$ unitary group of matrices
to represent the average signal-to-noise-ratio (SNR) optimal transmission and receive strategies along with the optimal phase adjustments at the RIS over Rayleigh fading. We prove that the left and right singular spaces of the RIS-receiver and transmit-RIS channel matrices are optimal in the above sense. Capitalizing on this, we exploit results from finite dimensional random matrix theory (see e.g., \cite{Chamain2020it}) to characterize the outage and throughput with and without the phase adjustment of the tiles of the RIS for each of the latter strategies. {Moreover, numerical results show that the proposed (suboptimal) strategy achieves comparable performance as the alternating  optimal strategy based on jointly optimized {\it instantaneous} SNR.
In particular, the throughput of the proposed scheme provides a tight lower bound for the  throughput corresponding to the latter strategy.}

%The most common joint optimization technique employed in the latter strategy, which is known as the alternating optimization method (a.k.a. block coordinate descent method) \cite{bertsekas1999nonlinear}, guarantees neither global nor local optimal solutions and requires centralized processing. However, the proposed new strategy does not suffer from those drawbacks. }  

%obtain exact expressions for the cumulative distribution functions (c.d.f.s) of the SNR with and without the phase adjustment of the tiles of the RIS for each of the latter strategies. We also derive the throughput as a consequence of the outage analysis. 
%Among other observations, our analytical results clearly show the outage and throughput improvement due to the phase adjustments of the tiles of the RIS. 

{\it Notations:} $\mathbb{E}\{\cdot\}$, $(\cdot)^\dagger$, $(\cdot)^T$, $||\cdot||$, $\text{Tr}(\cdot)$, $\text{diag}(\ldots)$, $|\cdot|$, $(\cdot)^*$, $\text{arg}(\cdot)$, and $\mathcal{U}_2$  denote the mathematical expectation, the Hermitian transpose operator, the conjugate operator, the transpose operator, the $\ell_2$ norm of a vector, the trace of a  matrix, a diagonal matrix, the modulus of a complex number, the complex conjugation operation, the argument of a complex number, and the group of $2\times 2$ unitary matrices, respectively. 

% {\it Notations:} $\mathbb{E}\{\cdot\}$ denotes the mathematical expectation, $(\cdot)^\dagger$ represents the Hermitian transpose operator, $||\cdot||$ stands for the $\ell_2$ norm of a vector. The trace of a square matrix is represented by $\text{Tr}(\cdot)$ and  $\text{diag}(\ldots)$ denotes a diagonal matrix, and $|\cdot|$ stands for the modulus of a complex number.

  \vspace{-2mm}

\section{System Model and Optimum Transmission Strategy} \label{S2}

% \begin{figure}
%   \centering
%   %\includegraphics[height=28mm,width=80mm]{SystemModel2.eps}\\
%   \includegraphics[width=0.35\textwidth]{system_model.eps}
%   \caption{Wireless communications via a two-element RIS.}\label{f_system_model_2}
% \end{figure}

We consider a wireless network in which a two-antenna transmitter and a two-antenna receiver communicate with the aid of an RIS. Based on \cite{najafi2020physicsbased}, we assume that the RIS is made of two continuous tiles whose surface phase shift can be appropriately optimized. The phase shift of each tile is the sum of a constant phase shift (denoted by $\phi_1$ and $\phi_2$) and a surface-dependent phase shift that depends on the point $(x,y)$ of the tile. The location-dependent phase-shifts are optimized in order to realize anomalous reflection based on the directions of the incident and reflected radio waves. The constant phase shifts, $\phi_1$ and $\phi_2$, are, on the other hand, optimized in order to maximize the combined SNR at the receiver. We focus our attention on the optimization of only $\phi_1$ and $\phi_2$, since the location-dependent phase-shifts are determined by the network geometry. {Further information on the optimization of the equivalent surface reflection coefficient  of anomalous reflectors can be found in \cite{danufane2020pathloss}.
} 
%The analysis of multiple tiles is postponed to future research. 

%A two-element RIS assisted one-way wireless network is considered with two-antenna transmitter and receiver. %, as shown in~Fig.~\ref{f_system_model_2}. 
The signal model can thus be written as
\begin{align}\label{e_system}
    \mathbf{y}=\mathbf{G}\boldsymbol{\Phi} \mathbf{H}\mathbf{x}+\mathbf{n}
\end{align}
where $\mathbf{G}\in\mathbb{C}^{2\times 1}$ is the RIS-to-receiver channel matrix, $\boldsymbol{\Phi}=\text{diag}\left(\exp(j\phi_1),\exp(j\phi_2)\right)\in\mathbb{C}^{2\times 2}$ with $\phi_i\in[-\pi,\pi)$ is the RIS reflection matrix which is diagonal by construction, $\mathbf{H}\in\mathbb{C}^{2\times 2}$ is the transmitter-to-RIS channel matrix, $\mathbf{x}\in \mathbb{C}^{2\times 1}$ is the transmit information vector and $\mathbf{n}\in\mathbb{C}^{2\times 2}\sim\mathcal{CN}_{2}\left(\mathbf{0}, \sigma^2\mathbf{I}_2\right)$ denotes the additive white Gaussian noise. %Moreover, we assume that 
Each element of $\mathbf{G}$ and $\mathbf{H}$ is distributed as $\mathcal{CN}(0,1)$ and $\mathbb{E}\left(\mathbf{x}\right)=\mathbf{0}$ and $\mathbb{E}\left(\mathbf{x}^\dagger \mathbf{x}\right)=\rho_s$. 
%It is worth noting that 
The channel matrices are referred to the equivalent channel from the transmitter to the tiles of the RIS and to the tiles of the RIS to the receiver. Rayleigh fading is assumed for analytical tractability and under the assumption that the location of the RIS cannot be optimized in order to ensure line-of-sight propagation \cite{qian2020beamforming}. {Moreover, we make the common assumption that  perfect channel state information (CSI) is available at the transmitter, RIS, and receiver. }
% The above matrix $\boldsymbol{\Phi}$ is commonly modelled as \cite{renzo2019corr,Wu2019twc,Han2019tvt,Nadeem2020twc,Jung2020twc,Basar2019acc,Zhao2020wcoml,Atapattu2020tcom,Kudathanthirige2020arxiv,Ferreira2020ojcomsoc,Badiu2020coml,qian2020beamforming,najafi2020physicsbased} $
%     \boldsymbol{\Phi}=\text{diag}\left(\exp(j\phi_1),\exp(j\phi_2)\right)
% $
% where $\phi_i\in[0,2\pi)$. 

By using the singular value decomposition, we rewrite \eqref{e_system} as
\[
    \mathbf{y}= \mathbf{U}\sqrt{\boldsymbol{\Lambda}}\mathbf{V}^\dagger\boldsymbol{\Phi} \mathbf{W}\sqrt{\boldsymbol{\Omega}}\mathbf{Q}^\dagger\mathbf{x}+\mathbf{n}
\]
where 
$\mathbf{G}=\mathbf{U}\sqrt{\boldsymbol{\Lambda}}\mathbf{V}^\dagger$, $\mathbf{H}=\mathbf{W}\sqrt{\boldsymbol{\Omega}}\mathbf{Q}^\dagger$,$\mathbf{U},\mathbf{V},\mathbf{W},\mathbf{Q}\in\mathcal{U}_2$, $\sqrt{\boldsymbol{\Lambda}}=\text{diag}\left(\sqrt{\lambda_1}, \sqrt{\lambda_2}\right)$, $\sqrt{\boldsymbol{\Omega}}=\text{diag}\left(\sqrt{\omega_1}, \sqrt{\omega_2}\right)$ with $\lambda_1>\lambda_2>0$ and $\omega_1>\omega_2>0$. In particular, $\lambda_1, \lambda_2$ and $\omega_1,\omega_2$ are the squares of the ordered singular values of $\mathbf{G}$ and $\mathbf{H}$ (i.e., ordered eigenvalues of $\mathbf{G}^\dagger\mathbf{G}$ and $\mathbf{H}^\dagger\mathbf{H}$), respectively. 
Let $\mathbf{a}, \mathbf{b}$ be the transmitting weight vector and receiver combining vector chosen such that $\mathbf{x}=\mathbf{a}s$ with $\mathbb{E}(s)=0$,
$\mathbb{E}(|s|^2)=\rho_s$, and $||\mathbf{b}||^2=1$. This in turn gives 
\[\mathbf{b}^\dagger \mathbf{y}=\sum_{k=1}^2\sum_{\ell=1}^2\sqrt{\lambda_k \omega_{\ell}} \mathbf{b}^\dagger\mathbf{u}_k\mathbf{v}_k^\dagger \boldsymbol{\Phi} \mathbf{w}_\ell\mathbf{q}_\ell^\dagger\mathbf{a}s+\mathbf{b}^\dagger\mathbf{n},\]
%To facilitate further analysis, we rewrite the above equation as
%\begin{align}
%\label{svdrearrang}
   % \mathbf{y}=\sum_{k=1}^2\sum_{\ell=1}^2\sqrt{\lambda_k \omega_{\ell}} \mathbf{u}_k\mathbf{v}_k^\dagger \boldsymbol{\Phi} \mathbf{w}_\ell\mathbf{q}_\ell^\dagger\mathbf{x}+\mathbf{n}.
%\end{align}
%\begin{align}
   % \mathbf{y}=\left(\sum_{k=1}^2\sqrt{\lambda_k} \mathbf{u}_k\mathbf{v}_k^\dagger\right) \boldsymbol{\Phi}\left(\sum_{\ell=1}^2\sqrt{\omega_\ell} \mathbf{w}_\ell\mathbf{q}_\ell^\dagger\right)\mathbf{x}+\mathbf{n}
%\end{align}
where we have used the column decompositions  $\mathbf{U}=\left(\mathbf{u}_1\; \mathbf{u}_2\right)$, $\mathbf{V}=\left(\mathbf{v}_1\; \mathbf{v}_2\right)$, $\mathbf{W}=\left(\mathbf{w}_1\; \mathbf{w}_2\right)$, and $\mathbf{Q}=\left(\mathbf{q}_1\; \mathbf{q}_2\right)$.
 %Now we may rearrange the terms in the above equation to yield
%\begin{align}
%\label{svdrearrang}
    %\mathbf{y}=\sum_{k=1}^2\sum_{\ell=1}^2\sqrt{\lambda_k \omega_{\ell}} \mathbf{u}_k\mathbf{v}_k^\dagger \boldsymbol{\Phi} \mathbf{w}_\ell\mathbf{q}_\ell^\dagger\mathbf{x}+\mathbf{n}.
%\end{align}
%Let $\mathbf{a}, \mathbf{b}$ be the transmitting weight vector and receiver combining vector chosen such that $\mathbf{x}=\mathbf{a}s$ with $\mathbb{E}(s)=0$,
%$\mathbb{E}(|s|^2)=\rho_s$, and $||\mathbf{b}||^2=1$\footnote{This normalization is assumed without loss of generality.}. This in turn gives
%\begin{align}
%\label{txweight}
   % \mathbf{y}=\sum_{k=1}^2\sum_{\ell=1}^2\sqrt{\lambda_k \omega_{\ell}} \mathbf{u}_k\mathbf{v}_k^\dagger \boldsymbol{\Phi} \mathbf{w}_\ell\mathbf{q}_\ell^\dagger\mathbf{a}s+\mathbf{n}.
%\end{align}
%\begin{align}\end{align}
Therefore, the instantaneous SNR at the receiver can be formulated as
\begin{align}
    \label{eq snr}
    \gamma=\bar{\gamma}\left|\sum_{k=1}^2\sum_{\ell=1}^2\sqrt{\lambda_k \omega_{\ell}} \mathbf{b}^\dagger\mathbf{u}_k\mathbf{v}_k^\dagger \boldsymbol{\Phi} \mathbf{w}_\ell\mathbf{q}_\ell^\dagger\mathbf{a}\right|^2
\end{align}
where $\bar{\gamma}=\rho_s/\sigma^2$. The following theorem gives the optimum transmission direction and combining vectors that maximize the average SNR in \eqref{eq snr}, for arbitrary phase shifts at the RIS.

\begin{theorem}
\label{thm main tx}
The transmission direction $\mathbf{a}=\mathbf{q}_1$ (i.e., leading right singular vector of $\mathbf{H}$) and the combining vector $\mathbf{b}=\mathbf{u}_1$ (i.e., leading left singular vector of $\mathbf{G}$) maximize the average SNR, whereas $\mathbf{a}=\mathbf{q}_2$ and $\mathbf{b}=\mathbf{u}_2$ minimize the average SNR. In particular, if we denote the SNR corresponding to the transmission direction $\mathbf{a}=\mathbf{q}_i$ and the combining vector $\mathbf{b}=\mathbf{u}_j$ by
$
    \gamma_j^{(i)}=\bar{\gamma} \lambda_j\omega_i Z_j^{(i)}
$, where $Z_j^{(i)}=\left|\mathbf{v}_j^\dagger \boldsymbol{\Phi}\mathbf{w}_i\right|^2$,
then %we have
\begin{align}
\label{eq snr order}
    \mathbb{E}\left\{ \gamma_{1}^{(1)}\right\}>\mathbb{E}\left\{\gamma_{2}^{(1)}\right\}&=\mathbb{E}\left\{\gamma_{1}^{(2)}\right\}>\mathbb{E}\left\{\gamma_{2}^{(2)}\right\}.
\end{align}

\end{theorem}
\begin{IEEEproof}
See Appendix \ref{app b sec a}.
\end{IEEEproof}

{\begin{remark}. The optimal transmission directions and combining vectors in Theorem~\ref{thm main tx} are independent of the phase shifts of the RIS, since the average SNR is optimized.
%based on the assumption that $\bf \Phi$ is functionally independent of $\bf H$ and $\bf G$.
This follows because, as stated in Appendix~\ref{app b sec a}, the distributions of $\mathbf{w}_i$ and $\boldsymbol{\Phi}\mathbf{w}_i$ are the same, under the assumption that $\bf \Phi$ is a constant matrix independent of the channels $\bf H$ and $\bf G$.
\end{remark}
%\new{
\begin{remark}
It is noteworthy that the  average SNR optimality of the right singular basis of $\mathbf{H}$ and the left singular basis of $\mathbf{G}$ in \eqref{eq snr order} remains true for an $n$-tile RIS assisted $n\times n$ MIMO system as well. 
\end{remark}}

Theorem~\ref{thm main tx} does not specify the exact choice of $\boldsymbol{\Phi}$ for each realization of the SNR, i.e., in the sense of maximizing the \textit{instantaneous} SNR.
%establishes the optimality of the right singular space of $\mathbf{H}$ and the left singular space of $\mathbf{G}$ in the sense of maximizing the {\it average} SNR. 
% However, it does not specify the exact choice of $\boldsymbol{\Phi}$ for each realization of the SNR, i.e., in the sense of maximizing the \textit{instantaneous} SNR. 
 To this end, using the notation $\mathbf{v}_j=(v_{j1}\; v_{j2})^T$ and $\mathbf{w}_i=(w_{i1}\; w_{i2})^T$, with the aid of the triangular inequality for the sum of two complex numbers, we obtain
\begin{align*}
    \gamma_j^{(i)}=\bar{\gamma} \lambda_j\omega_i \left|\mathbf{v}_j^\dagger \boldsymbol{\Phi}\mathbf{w}_i\right|^2\leq 
    \bar{\gamma}\lambda_j\omega_i
    \left(|v_{j1}||w_{i1}|+|v_{j2}||w_{i2}|\right)^2
\end{align*}
where the equality is achieved for 
\begin{align*}
%\label{eq phcomp}
\boldsymbol{\Phi}_{j, \text{cmp}}^{(i)}=\text{diag}\left(\exp\left\{-j\arg(v_{j1}^*w_{i1})\right\}, \exp\left\{-j\arg(v_{j2}^*w_{i2})\right\}\right)
.\end{align*}
{Therefore, since perfect CSI is available at the RIS, we can design the phases $\phi_1$ and $\phi_2$ as 
\begin{align}\label{eq_4}
    \phi_1=-\arg(v_{j1}^*w_{i1})\;\;\; \text{and}\;\;\; \phi_2=-\arg(v_{j2}^*w_{i2}).
\end{align}}
Accordingly, the phase-compensated SNR can be written as
\begin{align}
\label{eq snr compen}
     \gamma_{j,\text{cmp}}^{(i)}=\bar{\gamma}\lambda_j\omega_i  Z_{j,\sf{cmp}}^{(i)}
     \end{align}
     where 
     \begin{align*}
         Z_{j,\sf{cmp}}^{(i)}=
    \left(|v_{j1}||w_{i1}|+|v_{j2}||w_{i2}|\right)^2
.\end{align*}
Consequently, for $i,j=1,2$,
we can obtain the stochastic ordering relationship between the SNR before (i.e.,$ \gamma_{j}^{(i)}$ ) and after the compensation (i.e.,$\gamma_{j,\sf{cmp}}^{(i)}$ ) as
\begin{align*}
    \gamma_{j}^{(i)}<\gamma_{j,\sf{cmp}}^{(i)}
.
\end{align*}
Moreover, as shown in Corollary~\ref{cor avg comp order}, the average values of  $\gamma_{j,\sf{cmp}}^{(i)}$  follow the same ordering as in (\ref{eq snr order}).
\begin{corollary} \label{cor avg comp order}
The average values of the instantaneous SNR  after the phase-compensation, $\gamma_{j,{\sf cmp}}^{(i)}$,   satisfy the inequalities
\begin{align}
\label{eq avg comp ord}
   \hspace{-1mm}\mathbb{E}\left\{ \gamma_{1,{\sf cmp}}^{(1)}\right\}>\mathbb{E}\left\{\gamma_{2,{\sf cmp}}^{(1)}\right\}=\mathbb{E}\left\{\gamma_{1,
   \text{\sf cmp}}^{(2)}\right\}>\mathbb{E}\left\{\gamma_{2,\text{\sf cmp}}^{(2)}\right\}.
\end{align}
\end{corollary}
\begin{IEEEproof}
See Appendix \ref{app b sec b}.
\end{IEEEproof}

{\begin{remark}  It is worth pointing out that the compensated SNR in \eqref{eq snr compen} assumes that the transmission directions and the combining vectors are first optimized as  stated in Theorem~\ref{thm main tx}, and then the phase shifts of the RIS are optimized as in \eqref{eq_4}. In other words, these quantities are not jointly optimized.
%the transmission directions, the combining vectors, and the RIS phase shifts are not jointly optimized.
This results, in general, is a sub-optimal design, which has the advantage of providing analytical expressions for the transmission directions, the combining vectors, and the RIS phase shifts that are useful for analytical performance evaluations~\cite{zappone2020overheadaware}.
%The comparison with benchmark (based on numerical optimization) schemes is discussed in Section~\ref{s_num}.
\end{remark}}

Having established the  optimal properties of the proposed singular vector based transmission scheme, we are ready to analyze its performance in the following section.

  \vspace{-3mm}
%=========================
\section{Performance Analysis} \label{S3}
%=========================
We  study the effect of  phase adjustments at the RIS by deriving novel expressions for outage probability and throughput with and without the phase compensation at the RIS.

By definition, the outage probability can be written as \begin{align*}
    P^{\text{out}}(\gamma_{\text{th}})=\Pr\left\{\gamma\leq \gamma_{\text{th}}\right\}\end{align*}
where $\gamma\in \left\{\gamma_{j,\text{cmp}}^{(i)}, \gamma_{j}^{(i)}\right\}$ and $\gamma_{\text{th}}$ is an SNR threshold that is chosen according to the desired quality of service. Since we have already established the stochastic order $\gamma_{j,\text{cmp}}^{(i)}>\gamma_j^{(i)}$, it is not difficult to infer that, for $i,j=1,2$,
$
    P_{j,\text{cmp}}^{(i)}(\gamma_{\text{th}})<P_{j}^{(i)}(\gamma_{\text{th}}).
$
This in turn verifies our intuition that the phase compensation improves the system outage performance.
In order to compute outage probability, we first statistically characterize the random quantities $Z_{j,\text{cmp}}^{(i)}$ and $ Z_j^{(i)}$ by deriving closed-form expressions for their cumulative distribution functions (c.d.f.s), which are given by the following theorem.
\begin{theorem}\label{Theroem: CDF Z}
The c.d.f.s corresponding to  $Z_{j,\sf{cmp}}^{(i)}$ and $ Z_j^{(i)}$ are given, for $i,j=1,2,$ and $z\in(0,1)$, by
\begin{align}
    F_{Z_{j,\sf{cmp}}^{(i)}}(z)&=z-\sqrt{z(1-z)}\arcsin\left(\sqrt{z}\right)\label{thm2first},\\
      F_{Z_{j}^{(i)}}(z)&=z.\label{thm2second}
\end{align}
\end{theorem}
\begin{IEEEproof}
See Appendix \ref{app c}.
\end{IEEEproof}
\begin{remark}
It can be proved that, for an $n\times n$ RIS assisted MIMO system, $F_{Z_{j}^{(i)}}(z)$ takes the form $1-(1-z)^{n-1}$. 
\end{remark}
Consequently, we can evaluate the average SNR gain achieved by introducing the phase compensation at the RIS. To this end, we define the SNR gain as
\begin{align*}    \eta_{\text{Gain}}&=\displaystyle\frac{\mathbb{E}\left\{\gamma_{j,\text{cmp}}^{(i)}\right\}}{\mathbb{E}\left\{\gamma_{j}^{(i)}\right\}}
= \frac{\mathbb{E}\left\{Z_{j,\text{cmp}}^{(i)}\right\}}{\mathbb{E}\left\{Z_{j}^{(i)}\right\}}
    =1+\frac{\pi^2}{16}\approx 2\; {\rm dB}.
    \end{align*}
    This observation further strengthen the utility of phase compensation at the RIS.
    On the other hand, with or without phase compensation at the RIS, the average SNR gap (in dB) between any two consecutive transmission strategies is 
    \begin{align*}
        10\log\left(\mathbb{E}\left\{\gamma^{(1)}_j-\gamma_{2}^{(j)}\right\}\right)&=10\log\left(\mathbb{E}\left\{\gamma^{(1)}_{j,\text{cmp}}-\gamma_{2, \text{cmp}}^{(j)}\right\}\right)\\
        &=10\log\left(\frac{\mathbb{E}\left\{\omega_1\right\}}{\mathbb{E}\left\{\lambda_2\right\}}\right)=10\log 6\approx 7.8\;{\rm dB}
        \end{align*}
    where $j=1,2$. Therefore, we conclude that the phase compensation does not improve the average SNR gain between any two consecutive transmission strategies.

Armed with the above result,  we are in a position to derive  new  expressions of the outage probability. First, we focus on the outage corresponding to $\gamma_{j,\text{cmp}}^{(i)}$ given by 
\begin{align*}
P_{j,\text{cmp}}^{(i)}\left(\gamma_{\text{th}}/\bar{\gamma}\right) =\Pr\left\{\lambda_j\omega_i Z_{j,\text{cmp}}^{(i)}<\gamma_{\text{th}}/\bar{\gamma}\right\}. 
\end{align*}
Capitalizing on the independence between the singular values and singular vectors \cite{Shen2001} along with the independence of $\mathbf{G}$ and $\mathbf{H}$, we can rewrite the  outage probability as
\begin{align*}
P_{j,\text{cmp}}^{(i)}\left(\gamma_{\text{th}}/\bar{\gamma}\right)&= \mathbb{E}\left\{\Pr\left(\left.\lambda_j\leq \frac{\gamma_{\text{th}}/\bar{\gamma}}{\omega_i Z_{j,\text{cmp}}^{(i)}}\right|\omega_i,Z_{j,\text{cmp}}^{(i)}\right)\right\}
\end{align*}
where \cite{James1964ams}
\begin{align*}
F_{\lambda_1}(y)&=1-2\exp(-y)-y^2\exp(-y)+\exp(-2y),\\
F_{\lambda_2}(y)&=1-\exp(-2y),
\end{align*}
and the expected value is taken with respect to $f_{\omega_i}(x)=\displaystyle \frac{{\rm d}F_{\lambda_i}(x)}{{\rm d}x}$ 
%\begin{align*}
  %  &f_{\omega_i}(x)\\
   % &=\left\{
   % \begin{array}{ll}
   % \exp(-x)\left(2-2x+x^2-2\exp(-x)\right) & \text{for $i=1$}\\
    %2\exp(-2x) & \text{for $i=2$} .
   % \end{array}
   % \right.
%\end{align*}
and $f_{Z_{j,\text{cmp}}^{(i)}}(y)=\displaystyle\frac{{\rm d} F_{Z_{j,\text{cmp}}^{(i)}}(y)}{{\rm d}y}$.
% \begin{align*}
%     &F_{\lambda_j}(y)\\
%     &=\left\{
%     \begin{array}{ll}
%     1-2\exp(-y)-y^2\exp(-y)+\exp(-2y) & \text{for $j=1$}\\
%     1-\exp(-2y) & \text{for $j=2$} .
%     \end{array}
%     \right.
% \end{align*}
Since the eigenvalues $\omega_i$ and $\lambda_i$, for $i=1,2$, have identical distributions (i.e., because $\mathbf{G}$ and $\mathbf{H}$ have identical distributions), we can readily obtain the relation, $P_{1,\text{cmp}}^{(2)}\left(\gamma_{\text{th}}/\bar{\gamma}\right)=P_{2,\text{cmp}}^{(1)}\left(\gamma_{\text{th}}/\bar{\gamma}\right)$. Moreover, a similar approach with $f_{Z_{j}^{(i)}}(y)=1$, for $y\in(0,1)$, can be used to derive the outage probability corresponding to $\gamma_{j}^{(i)}$ given by 
\begin{align*}P_{j}^{(i)}\left(\gamma_{\text{th}}/\bar{\gamma}\right)=\mathbb{E}\left\{\Pr\left(\left.\lambda_j\leq \frac{\gamma_{\text{th}}/\bar{\gamma}}{\omega_i Z_{j}^{(i)}}\right|\omega_i,Z_{j}^{(i)}\right)\right\}.
\end{align*}
Finally, some  algebraic manipulations give the corresponding outage expressions as shown in the following corollary.
\begin{corollary}\label{cor outage}
The outage probabilities $P_{j,\text{cmp}}^{(i)}\left(\gamma_{\text{th}}/\bar{\gamma}\right)$ and $P_{j}^{(i)}\left(\gamma_{\text{th}}/\bar{\gamma}\right)$ can be formulated as 
\begin{align}
   P_{1,\text{cmp}}^{(1)}(z)&=1-4\mathcal{I}_0(1,1,z)+4\mathcal{I}_0(2,1,z)-2\mathcal{I}_0(3,1,z)+4\mathcal{I}_0(1,2,z) -2z^2\mathcal{I}_2(-1,1,z)\nonumber\\ 
   & \qquad \qquad+2z^2\mathcal{I}_2(0,1,z)
    -z^2\mathcal{I}_2(1,1,z)+2z^2\mathcal{I}_2(-1,2,z) +2\mathcal{I}_0(1,1,2z) \nonumber\\ 
   & \qquad \qquad 
   -2\mathcal{I}_0(2,1,2z)+\mathcal{I}_0
    (3,1,2z)-2\mathcal{I}_0(1,2,2z)\label{maxcmpoutage}\\
     P_{2,\text{cmp}}^{(1)}(z)&=
     P_{1,\text{cmp}}^{(2)}(z)=1-4\mathcal{I}_0(1,2,z)-2z^2\mathcal{I}_2(-1,2,z)+2\mathcal{I}_0(1,2,2z)
    \label{midcmpoutage}\\
   P_{2,\text{cmp}}^{(2)}(z)&=1-2\mathcal{I}_0(1,2,2z)\label{mincmpoutage}\\
   P_{1}^{(1)}(z)&=1
    -4z^2\mathcal{G}(z,1)
    +8\sqrt{z}K_1(2\sqrt{z})
    -2z^2\mathcal{G}(z,-1)
    +8z^2 \mathcal{G}(2z,1)
    -4z K_0(2\sqrt{z})\nonumber\\
    &\qquad \qquad +4z\sqrt{z}K_1(2\sqrt{z}) -2z^2 K_2(2\sqrt{z})+4z K_0(2\sqrt{2z})+8z^2 \mathcal{G}(2z,1) \nonumber\\
    &\qquad \qquad -4\sqrt{2z}K_1(2\sqrt{2z})+4z^2 \mathcal{G}(2z,-1)-16z^2 \mathcal{G}(4z,1)
    \label{maxuncoutage}\\
     P_{2}^{(1)}(z)&= P_{1}^{(2)}(z)=
     1-8z^2 \mathcal{G}(2z,1)-4zK_0(\sqrt{8z})+16z^2 \mathcal{G}(4z,1)
     \label{miduncoutage}\\
    P_{2}^{(2)}(z)&=1-16z^2\mathcal{G}(4z,1) \label{minuncoutage}
\end{align}
where 
$\mathcal{G}(z,a)=G_{1,3}^{3,0} \left(z\bigg |\begin{matrix}0 \\                  
-1, -a,-2
                          \end{matrix} \right)$ denotes the Meijer G-function,
                          \begin{align*}
              \mathcal{I}_a(\alpha,\gamma,\beta z)=&\frac{\left(\beta z\right)^{2-a}}{2\gamma^{\alpha+a-2}}
        G_{1,3}^{3,0} \left(\beta \gamma z\bigg |\begin{matrix}0 \\                  
-1, \alpha+a-2,a-2
                          \end{matrix}               \right) \\  & \qquad \qquad \qquad +\left(\frac{\beta z}{\gamma}\right)^{\frac{\alpha}{2}} \int_1^\infty t^{a+\frac{\alpha}{2}-2}\frac{(2-t)}{\sqrt{t-1}} K_{\alpha}\left(2\sqrt{\beta \gamma z t}\right)\arcsin \left(\frac{1}{\sqrt{t}}\right) {\rm d}t
                          \end{align*}
%               \[\mathcal{I}_a(\alpha,\gamma,\beta z)=\frac{\left(\beta z\right)^{2-a}}{2\gamma^{\alpha+a-2}}
%         G_{1,3}^{3,0} \left(\beta \gamma z\bigg |\begin{matrix}0 \\                  
% -1, \alpha+a-2,a-2
%                           \end{matrix}               \right)+\left(\frac{\beta z}{\gamma}\right)^{\frac{\alpha}{2}} \int_1^\infty t^{a+\frac{\alpha}{2}-2}\frac{(2-t)}{\sqrt{t-1}} K_{\alpha}\left(2\sqrt{\beta \gamma z t}\right)\arcsin \left(\frac{1}{\sqrt{t}}\right) {\rm d}t\]            
    %\begin{align}
   % \label{Ifunction}
       % \mathcal{I}_a(\alpha,\gamma,\beta z)&=\frac{\left(\beta z\right)^{2-a}}{2\gamma^{\alpha+a-2}}
       % G_{1,3}^{3,0} \left(\beta \gamma z\bigg |\begin{matrix}0 \\                  
%-1, \alpha+a-2,a-2
                        %  \end{matrix}                         \right)\nonumber\\
                         % &+\left(\frac{\beta z}{\gamma}\right)^{\alpha/2} \int_1^\infty t^{a+\alpha/2-2}(2-t)(t-1)^{-1/2}\nonumber\\
                         % &\qquad \qquad \quad \times K_{\alpha}\left(2\sqrt{\beta \gamma z t}\right)\arcsin \left(\frac{1}{\sqrt{t}}\right) {\rm d}t,
   % \end{align}
        and $K_\alpha(z)$ is the modified Bessel function of the second kind and order $\alpha$.
\end{corollary}

To demonstrate the utility of the newly derived  expressions, we can  write the average throughput (in nats/sec/Hz) of each of the phase compensated transmission strategies, for $i,j=1,2$, as \[R_{j,\text{cmp}}^{(i)}=
    \int_0^\infty \frac{1-P_{j,\text{cmp}}^{(i)}(z/\bar{\gamma})}{1+z} {\rm d}z,\] 
% \begin{align*}
%     R_{j,\text{cmp}}^{(i)}=\mathbb{E}\{\ln(1+\gamma_{j,\text{cmp}}^{(i)})\}=
%     \int_0^\infty \frac{1-P_{j,\text{cmp}}^{(i)}(z/\bar{\gamma})}{1+z} {\rm d}z,
% \end{align*}
whereas the results corresponding to the uncompensated phases are given by \[R_{j}^{(i)}=\int_0^\infty \frac{1-P_{j}^{(i)}(z/\bar{\gamma})}{1+z} {\rm d}z.\] To compute the average throughput, we encounter integrals of the form $\int_0^\infty z^m \mathcal{I}_a(\alpha, \gamma, \beta z/\bar{\gamma})/(1+z) {\rm d}z$, which can be solved using \cite[Eq.~7.811.5]{gradshteyn2007book} to obtain an expression involving a single integral. For instance, the above procedure gives
\[R_{2,\text{cmp}}^{(2)}{=}\frac{1}{2}
                          \int_1^\infty \frac{(2-t)}{t^2 \sqrt{t-1}} \arcsin \left(\frac{1}{\sqrt{t}}\right) G_{1,3}^{3,1} \left(\frac{4t}{\bar{\gamma}}\bigg |\begin{matrix}0 \\                  
0,1,0
                          \end{matrix} \right) {\rm d}t\\
                           +
                          \frac{1}{2}R_{2}^{(2)}\]
% \begin{align*}
%      R_{2,\text{cmp}}^{(2)}{=}\frac{1}{2}
%                           \int_1^\infty \frac{(2-t)}{t^2 \sqrt{t-1}} &\arcsin \left(\frac{1}{\sqrt{t}}\right) G_{1,3}^{3,1} \left(\frac{4t}{\bar{\gamma}}\bigg |\begin{matrix}0 \\                  
% 0,1,0
%                           \end{matrix} \right) {\rm d}t\\
%                           & +
%                           \frac{8}{\bar{\gamma}^2}
%      G_{2,4}^{4,1} \left(\frac{4}{\bar{\gamma}}\bigg |\begin{matrix}-2,0 \\                  
% -2, -1,-1,-2
%                           \end{matrix} \right)
% \end{align*}
where \[R_{2}^{(2)}=\displaystyle \frac{16}{\bar{\gamma}^2} 
     G_{2,4}^{4,1} \left(\frac{4}{\bar{\gamma}}\bigg |\begin{matrix}-2,0 \\                  
-2, -1,-1,-2
                          \end{matrix} \right).\] %It is noteworthy that the above integral can easily be evaluated numerically for all $\bar{\gamma}$.
                         
\section{Numerical Results}\label{s_num}
\begin{figure}[!t]
  \vspace{-1mm}
  \centering
  \includegraphics[width=0.7\textwidth]{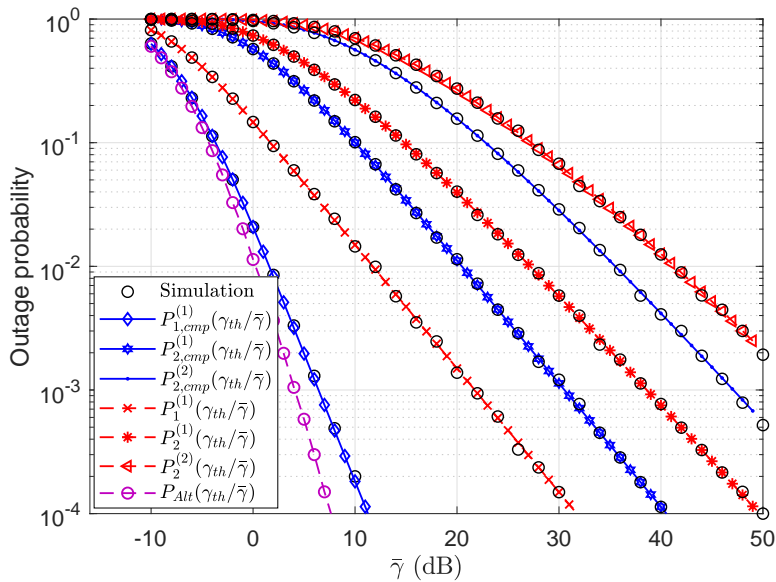}\\
  \caption{Outage probability vs average SNR for $\gamma_{\text{th}}=0$\,dB.
  }\label{f_out}
  \vspace{-3mm}
\end{figure}

   In this section, we present some numerical results to illustrate the accuracy of our analysis and to establish the optimality of the proposed transmission scheme. To this end, let us first establish the accuracy of the outage expressions given in (\ref{maxcmpoutage})-(\ref{minuncoutage}).  Fig. \ref{f_out} compares the outage probability as a function of $\bar{\gamma}$ (in dB) for a fixed threshold $\gamma_{\text{th}}=0\; {\rm dB}$ in the presence and in the absence of an RIS. The analytical curves generated based on Corollary \ref{cor outage} match precisely with the simulations. Moreover, as expected, the phase compensation at the RIS leads to an outage improvement for all $\bar{\gamma}$. {On the other hand, to establish the optimality of the compensated SNR in \eqref{eq snr compen}, we maximize $\gamma=\bar{\gamma}\left|\mathbf{b}^\dagger \mathbf{G}\boldsymbol{\Phi}\mathbf{H}\mathbf{a}\right|^2$ subject to $||\mathbf{a}||=1, ||\mathbf{b}||=1$, and $\boldsymbol{\Phi}\boldsymbol{\Phi}^*=\mathbf{I}_2$ and compare the resultant outage expression with $P_{1,\text{cmp}}^{(1)}(\gamma_{\text{th}}/\bar{\gamma})$. Since this problem is non-convex, we exploit the alternating optimization procedure detailed in \cite{bertsekas1999nonlinear} to obtain the optimal $\gamma$. Having noted that this procedure is sensitive to the initial condition, we initialize with $\boldsymbol{\Phi}=\boldsymbol{\Phi}_{1,\text{cmp}}^{(1)}$ and solve the sub problem $\max\bar{\gamma}\left|\mathbf{b}^\dagger \mathbf{G}\boldsymbol{\Phi}_0\mathbf{H}\mathbf{a}\right|^2$ subject to $||\mathbf{a}||=1$ and $||\mathbf{b}||=1$ to obtain the optimal $\mathbf{a}_0$ and $\mathbf{b}_0$. Subsequently we solve the sub problem $\max\bar{\gamma}\left|\mathbf{b}_0^\dagger \mathbf{G}\boldsymbol{\Phi}\mathbf{H}\mathbf{a}_0\right|^2$ subject to  $\boldsymbol{\Phi}\boldsymbol{\Phi}^*=\mathbf{I}_2$ to determine the optimal $\boldsymbol{\Phi}_1$. We continue these iterations till convergence to obtain the numerically optimal $\mathbf{a}, \mathbf{b}$, and $\boldsymbol{\Phi}$. If we denote the corresponding optimal instantaneous SNR as $\gamma_{\text{Alt}}$, then we have the stochastic ordering $\gamma_{\text{Alt}}\geq \gamma_{1,\text{cmp}}^{(1)}\geq \gamma_1^{(1)}$ from which we obtain
   \begin{align*}
       P_{\text{Alt}}(\gamma_{\text{th}}/\bar{\gamma})
       \leq 
       P_{1,\sf{cmp}}^{(1)}(\gamma_{\text{th}}/\bar{\gamma})
       \leq
       P_{1}^{(1)}(\gamma_{\text{th}}/\bar{\gamma}).
   \end{align*}
   %Since we can terminate the iterations by solving an arbitrary sub problem upon convergence, after several updates, we can conveniently obtain the optimal SNR as
  % $\gamma_{\text{Alt}}=\bar{\gamma} \left|\mathbf{b}_K^\dagger\mathbf{G} \boldsymbol{\Phi}_K\mathbf{H}\mathbf{a}_K\right|^2=\bar{\gamma}\sigma^2_{\max}\left(\mathbf{G} \boldsymbol{\Phi}_K\mathbf{H}\right)=\bar{\gamma}\lambda_{\max}\left(\mathbf{G} \boldsymbol{\Phi}_K\mathbf{H} \mathbf{H}^\dagger \boldsymbol{\Phi}_K^\dagger \mathbf{G}^\dagger\right)$, where $\boldsymbol{\Phi}_K$ is the last updated $\boldsymbol{\Phi}$. Since $\lambda_{\max}\left(\mathbf{G} \boldsymbol{\Phi}_K\mathbf{H} \mathbf{H}^\dagger \boldsymbol{\Phi}_K^\dagger \mathbf{G}^\dagger\right)$ and $\lambda_{\max}\left(\mathbf{G} \mathbf{H} \mathbf{H}^\dagger \mathbf{G}^\dagger\right)$ have the same distribution, the corresponding outage probability takes the form $ P_{\text{Alt}}(\gamma_{\text{th}}/\bar{\gamma})=\Pr\left\{\gamma_{\text{Alt}}\leq \gamma_\text{th}\right\}=
       % \Pr\left\{\lambda_{\max}\left(\mathbf{G} \mathbf{H} \mathbf{H}^\dagger \mathbf{G}^\dagger\right)\leq 
       % \gamma_\text{th}/\bar{\gamma}\right\}$.
        For comparison, in Fig. 1, the outage corresponding to the above optimal scheme $P_{\text{Alt}}(\gamma_{\text{th}}/\bar{\gamma})$ is also plotted for $\gamma_{\text{th}}=0$ \text{dB}. As can be seen from the figure, although sub-optimal, the proposed scheme (i.e., $P_{1,\text{com}}^{(1)}(\gamma_{\text{th}}/\bar{\gamma})$) provides a good approximation to the optimal outage, particularly in the low SNR regime.}

   %Here in each outage expression, the superscript $i$ denotes the $i$th column of $\mathbf{Q}$  along which the data is transmitted and the subscript $j$ stands for the corresponding combining vector chosen from the $j$th column of $\mathbf{U}$.
   %Moreover, as expected, the phase compensation at the RIS leads to an outage improvement for all $\bar{\gamma}$. These results further strengthen the claim that the transmission along the leading left singular vector of $\mathbf{H}$ followed by combining using the right leading singular vector of $\mathbf{G}$ is optimal in the sense of average SNR. However, more interesting observation is that the latter transmission-receive directions have a very significant outage disparity with the other strategies, particularly in the high SNR regime (i.e., large $\bar{\gamma}$). Moreover, as can be seen from the figure, this particular transmission strategy achieves a significant diversity gain over the other strategies. 
   
\begin{figure}[!t]
  \vspace{-1mm}
  \centering
  \includegraphics[width=0.7\textwidth]{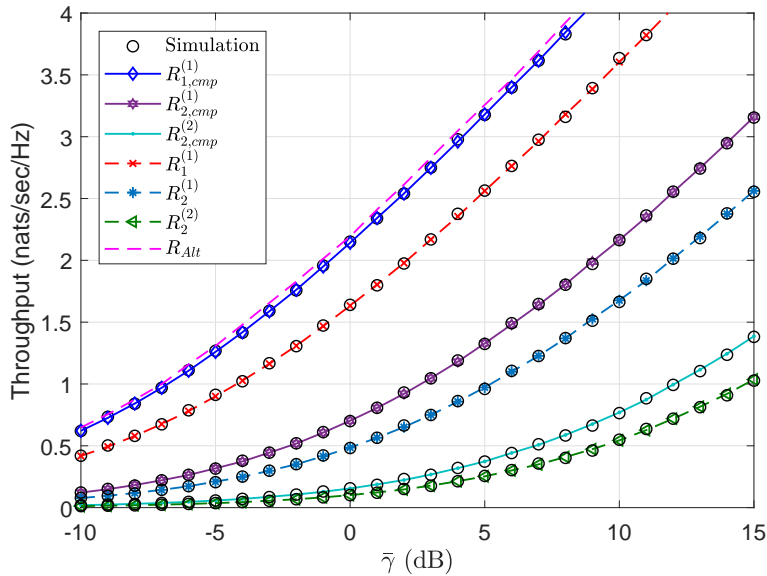}\\
  \caption{Average throughput vs average SNR.
  }\label{f_tp}
    \vspace{-3mm}
\end{figure}

Figure \ref{f_tp} compares the throughput in nats/sec/Hz as a function of $\bar{\gamma}$ (in dB) in the presence and absence of the RIS. A good agreement between the analytical and simulation results can be observed, and the improvement of the throughput due to the phase compensation at the RIS is clearly visible in the figure. {Again, for comparison, the throughput $R_{\text{Alt}}$ corresponding to the above numerically optimized scheme is also plotted. As depicted in the figure, the proposed scheme provides a very tight lower bound to the optimal throughput. This claim further highlights the utility of the proposed scheme.}

%Also, in the high SNR regime (i.e., large $\bar{\gamma}$ in dB), the throughput shows a linear trend as expected. Moreover, the high SNR power offset  is significantly low for the leading singular vector transmission-receive scheme in comparison with the other modes of transmission. 

%A. Lozano, A. M. Tulino, and S. Verdu ́, “High-SNR power offset in multiantenna communication,” IEEE Trans. Inform.
%Theory, vol. 51, no. 12, pp. 4134–4151, Dec. 2005.

\vspace{-2mm}   
\section{Conclusion}
We have analytically characterized the exact outage and throughput of a two-tile RIS-assisted $2\times 2$ wireless network in the presence of Rayleigh fading. In particular, we have considered transmission strategies corresponding to all combinations of   left  and  right  singular spaces of the RIS-receiver and transmit-RIS channels that are optimal in terms of the average SNR. {It turns out that, although sub-optimal, the proposed strategy  has outage and throughput performances comparable with the numerically jointly optimized scheme. 
In particular, with respect to throughput, the proposed scheme provides a tight lower bound to the numerically optimal scheme. 
Moreover, we have proved that the average SNR improves of about 2 dB thanks to the phase compensation introduced by the RIS.}  
%Moreover, our results clearly show the performance improvement due to the phase adjustment at the RIS.
%Moreover, we experience $2$\; dB average SNR improvement due the phase compensation at the RIS. 
%\appendix
\numberwithin{equation}{section}
\vspace{-2mm}  
\begin{appendices}
\section{}
To facilitate our main derivations, we will require the following preliminary result. 
\begin{lemma}\label{lem unitary character}
Let $\mathbf{S}\in \mathcal{U}_2$. Then $\mathbf{S}$ can be parameterized as \cite{Spengler2012}
\begin{align*}
\mathbf{S}=\left(\begin{array}{cr}
    e^{j\theta_{11}}\cos \theta_{12} & -e^{ j\left(\theta_{11}+\theta_{21}\right)}\sin \theta_{12}\\
    e^{ j \theta_{22}} \sin \theta_{12} & e^{ j\left(\theta_{22}+\theta_{21}\right)}\cos \theta_{12}
    \end{array}\right)
\end{align*}
where $\theta_{ij}\in [0,2\pi)$ for $i\geq j$ and $\theta_{12}\in[0,\pi/2]$. Moreover,
if $\mathbf{S}$ is uniformly distributed over $\mathcal{U}_2$, then all the entries of $\mathbf{S}$ are identically distributed and the normalized invariant measure ${\rm d}\mathbf{S}$ (i.e, Haar measure) defined on $\mathcal{U}_2$ such that $\int_{\mathcal{U}_2} {\rm d}\mathbf{S}=1$ is given by \cite{Spengler2012}
\begin{align*}
    \mathrm{d} \mathbf{S}=\frac{1}{8\pi^3}\sin2\theta_{12}{\rm d}\theta_{12}{\rm d}\theta_{21} {\rm d}\theta_{11}{\rm d}\theta_{22}.
\end{align*}
Therefore, the random variables $\theta_{ij}$ are independent for $i,j=1,2$ with their densities given by $
    f_{\theta_{12}}(\theta)=\sin 2\theta,\;\;\; 0\leq \theta \leq \pi/2,
$ and 
$
  \theta_{ij}\sim \text{Uniform}(0,2\pi),\;\;\; j\leq i. $
  \end{lemma}
\vspace{-3mm}   
\subsection{Proof of Theorem~\ref{thm main tx} } \label{app b sec a}
Since the column spaces of $\mathbf{U}$ and $\mathbf{Q}$ form orthonormal bases, we can find $\boldsymbol{\alpha}, \boldsymbol{\beta}\in\mathbb{C}^{2\times 1}$ such that
$
    \mathbf{b}=\mathbf{U}\boldsymbol{\alpha}
$
and 
$
    \mathbf{a}=\mathbf{Q} \boldsymbol{\beta}
$
with $||\boldsymbol{\alpha}||^2=||\boldsymbol{\beta}||^2=1$.
If we denote $\boldsymbol{\alpha}=\left(\alpha_1\; \alpha_2\right)^T$ and $\boldsymbol{\beta}=\left(\beta_1\; \beta_2\right)^T$, then (\ref{eq snr}) can be  written as 
\[    \gamma=\bar{\gamma} \sum_{k,\ell,i,j=1}^2
    \sqrt{\lambda_k \omega_{\ell}\lambda_i \omega_{j}}
    \alpha_k^* \alpha_i  \beta_l \beta_j^* \text{Tr}\left(\mathbf{v}_i\mathbf{v}_k^\dagger \boldsymbol{\Phi} \mathbf{w}_\ell \mathbf{w}_j^\dagger \boldsymbol{\Phi}^\dagger\right).\] 
% \begin{align*}
%     \gamma&=\bar{\gamma} \sum_{k,\ell,i,j=1}^2
%     \sqrt{\lambda_k \omega_{\ell}\lambda_i \omega_{j}}
%     \alpha_k^* \alpha_i  \beta_l \beta_j^* \text{Tr}\left(\mathbf{v}_i\mathbf{v}_k^\dagger \boldsymbol{\Phi} \mathbf{w}_\ell \mathbf{w}_j^\dagger \boldsymbol{\Phi}^\dagger\right).
% \end{align*}
Capitalizing on the independence between the singular values and singular vectors \cite{Shen2001} and 
noting that, under the assumption that $\bf \Phi$ is a constant matrix that is not optimized and is independent of the wireless channels, the distributions of $\boldsymbol{\Phi}\mathbf{w}_\ell$ and $\boldsymbol{\mathbf{w}_j^\dagger\Phi}^\dagger$ are the same as $\mathbf{w}_\ell$ and $\mathbf{w}_j$, we obtain
\begin{align}
\label{eq avg app}
   \mathbb{E}\left\{ \gamma\right\}&=\bar{\gamma} \sum_{k,\ell,i,j=1}^2\alpha_k^* \alpha_i  \beta_\ell \beta_j^*
    \mathbb{E}\left\{\sqrt{\lambda_k \omega_{\ell}\lambda_i \omega_{j}}\right\}
 \text{Tr}\left(\mathbb{E}\left\{\mathbf{v}_i\mathbf{v}_k^\dagger\right\} \mathbb{E}\left\{ \mathbf{w}_\ell \mathbf{w}_j^\dagger\right\}\right).
\end{align}
In order to evaluate the expected values inside the trace operator, following Lemma \ref{lem unitary character}, we parameterize the matrices $\mathbf{V}$ and $\mathbf{W}$  to yield
\[
    \mathbf{V}=\left(\begin{array}{cr}
    e^{j\xi_{11}}\cos \xi_{12} & -e^{ j\left(\xi_{11}+\xi_{21}\right)}\sin \xi_{12}\\
    e^{ j \xi_{22}} \sin \xi_{12} & e^{ j\left(\xi_{22}+\xi_{21}\right)}\cos \xi_{12}
    \end{array}\right),
\]
    and
    \[
        \mathbf{W}=\left(\begin{array}{cr}
    e^{j\psi_{11}}\cos \psi_{12} & -e^{ j\left(\psi_{11}+\psi_{21}\right)}\sin \psi_{12}\\
    e^{ j \psi_{22}} \sin \psi_{12} & e^{ j\left(\psi_{22}+\psi_{21}\right)}\cos \psi_{12}
    \end{array}\right),
    \]
    where $\xi_{ij},\psi_{ij}\in [0,2\pi)$ for $i\geq j$ and $\xi_{12},\psi_{12}\in[0,\pi/2]$. Moreover, since $\mathbf{V}$ and $\mathbf{W}$ are uniformly  and independently distributed over $\mathcal{U}_2$ \cite{Shen2001}, $\xi_{ij}$ and $\psi_{ij}$ are independent and identically distributed (i.i.d.)\cite{Spengler2012}.
 Therefore, we obtain
 \[
        \mathbb{E}\left\{\mathbf{v}_i\mathbf{v}_k^\dagger\right\}=\left\{\begin{array}{cc}
        \frac{1}{2}\mathbf{I}_2 & \text{if $i=k$}\\
        \mathbf{0} & \text{if $i\neq k$}
        \end{array}
        \right.
  \]
and 
\[
        \mathbb{E}\left\{\mathbf{w}_\ell\mathbf{w}_j^\dagger\right\}=\left\{\begin{array}{cc}
        \frac{1}{2}\mathbf{I}_2 & \text{if $j=\ell$}\\
        \mathbf{0} & \text{if $j\neq \ell$}
        \end{array}
        \right.
    .\]
Then, (\ref{eq avg app}) gives
\begin{align*}
    &\mathbb{E}\left\{ \gamma\right\}
    %\nonumber\\     &
    =\frac{\bar{\gamma}}{2}
    \boldsymbol{\alpha}^\dagger \left(\begin{array}{cc}
    \hspace{-1mm}\mathbb{E}\{\lambda_1\} & \hspace{-1mm}0\\
    \hspace{-1mm}0 & \hspace{-1mm}\mathbb{E}\{\lambda_2\}\end{array}\hspace{-2mm}\right) \boldsymbol{\alpha}
    \boldsymbol{\beta}^\dagger \left(\begin{array}{cc}
    \hspace{-1mm}\mathbb{E}\{\omega_1\} & \hspace{-1mm}0\\
    \hspace{-1mm}0 & \hspace{-1mm}\mathbb{E}\{\omega_2\}\end{array}\hspace{-2mm}\right) \boldsymbol{\beta}
\end{align*}
where $\mathbb{E}\{\lambda_1\}>\mathbb{E}\{\lambda_2\}$ and $\mathbb{E}\{\omega_1\}>\mathbb{E}\{\omega_2\}$.
Clearly, $\boldsymbol{\alpha}=e^{j\chi_1}\left(1\; 0\right)^T$ and $\boldsymbol{\beta}=e^{j\chi_2}\left(1\; 0\right)^T$, with $\chi_1,\chi_2\in[0,2\pi)$ denoting arbitrary phases, simultaneously maximize $\mathbb{E}\{\gamma\}$ subject to the constraints $||\boldsymbol{\alpha}||^2=||\boldsymbol{\beta}||^2=1$. Therefore,  we  conclude that the pair $(\mathbf{a},\mathbf{b})\equiv\left(\mathbf{q}_1,\mathbf{u}_1\right)$ is optimal in the sense of maximizing the average SNR. Similar arguments can be used to establish that $(\mathbf{a},\mathbf{b})\equiv(\mathbf{q}_2,\mathbf{u}_2)$ corresponds to the minimum average SNR. 

Having noted that $ \gamma_j^{(i)}=
%\bar{\gamma} \lambda_j\omega_i \left|\mathbf{v}_j^\dagger \boldsymbol{\Phi}\mathbf{w}_i\right|^2
\bar{\gamma} \lambda_j\omega_i \text{Tr}\left(\mathbf{v}_j\mathbf{v}_j^\dagger \boldsymbol{\Phi}\mathbf{w}_i \mathbf{w}_i^\dagger \boldsymbol{\Phi}^\dagger\right)$ and 
$
    \mathbb{E}\left\{\text{Tr}\left(\mathbf{v}_j\mathbf{v}_j^\dagger \boldsymbol{\Phi}\mathbf{w}_i \mathbf{w}_i^\dagger \boldsymbol{\Phi}^\dagger\right)\right\}
    %=\mathbb{E}\left\{\text{Tr}\left(\mathbf{v}_j\mathbf{v}_j^\dagger \mathbf{w}_i \mathbf{w}_i^\dagger \right)\right\}
    =\frac{1}{2},
$
we obtain $\mathbb{E}\left\{ \gamma_j^{(i)}\right\}=\frac{\bar{\gamma}}{2} \mathbb{E}\{\lambda_j\} \mathbb{E}\{\omega_i\}$. Finally, we use the inequalities
\begin{align}
   \label{eq eigorder}
       \lambda_1\omega_1>\lambda_1\omega_2>\lambda_2\omega_2 \quad \text{and} \quad
       \lambda_1\omega_1>\lambda_2\omega_1>\lambda_2\omega_2,
   \end{align}
along with the fact that $\mathbb{E}\{\lambda_1\}\mathbb{E}\{\omega_2\}=\mathbb{E}\{\lambda_2\}\mathbb{E}\{\omega_1\}$ to obtain (\ref{eq snr order}) which concludes the proof.

\subsection{Proof of Corollary \ref{cor avg comp order}}\label{app b sec b}
Following the above parametatrizations of unitary matrices, we can rewrite (\ref{eq snr compen}) as
$
   \gamma_{i, \text{comp}}^{(i)}=\bar{\gamma}\lambda_i\omega_i \cos^2\left(\xi_{12}-\psi_{12}\right), \;\; i=1,2 
$
and 
$
    \gamma_{j, \text{comp}}^{(i)}=\bar{\gamma}\lambda_j\omega_i \sin^2\left(\xi_{12}+\psi_{12}\right), \;\; i\neq j.
$

Since 
%$\xi_{12}$ and $\psi_{12}$ have identical distributions over $[0,\pi/2]$, it is not difficult to establish 
$
    \mathbb{E}\left\{\cos^2\left(\xi_{12}-\psi_{12}\right)\right\}=\mathbb{E}\left\{\sin^2\left(\xi_{12}+\psi_{12}\right)\right\},
    $
 we use the inequalities (\ref{eq eigorder}) and the relation $\mathbb{E}\{\lambda_1\}\mathbb{E}\{\omega_2\}=\mathbb{E}\{\lambda_2\}\mathbb{E}\{\omega_1\}$
   to obtain (\ref{eq avg comp ord}). This concludes the proof.

\vspace{-3mm}    
\subsection{Proof of Theorem~\ref{Theroem: CDF Z}}\label{app c}
    %We may use 
    The unitary matrix parameterizations given in (\ref{app b sec a}) yields
    \begin{align}
        Z_{j,\text{cmp}}^{(i)}=\left\{\begin{array}{cc}\cos^2\left(\xi_{12}-\psi_{12}\right) & \text{for $i=j$}\\
        \sin^2\left(\xi_{12}+\psi_{12}\right) & \text{for $i\neq j$}
        \end{array}\right.
    \end{align}
    where the distributions of $\xi_{12}$ and $\psi_{12}$ are i.i.d. with the common density
    $
        f(\theta)=\sin2\theta,\;\; \theta\in[0,\pi/2]
    $ \cite{Spengler2012}.
    Therefore, simple variable transformations yield \[f_{\left(\xi_{12}-\psi_{12}\right)}(x)=\frac{1}{2}\left(\frac{\pi}{2}\cos 2x-|x|\cos 2x+\frac{1}{2}\sin 2|x|\right),\;\; x\in[-\pi/2,\pi/2]\]  and
    \begin{align*}
        f_{\xi_{12}+\psi_{12}}(x)=\left\{\begin{array}{lc}
        \frac{1}{4}\sin 2x -\frac{x}{2}\cos 2x & \text{for $x\in\left[0,\frac{\pi}{2}\right)$}\\
        -\frac{1}{4}\sin 2x -\frac{(\pi -x)}{2}\cos 2x & \text{for $x\in\left[\frac{\pi}{2},\pi\right]$}.
        \end{array}\right.
    \end{align*}
    %\begin{align}
        %f_{\left(\xi_{12}-\psi_{12}\right)}(x)=\frac{1}{2}\left(\frac{\pi}{2}\cos 2x-|x|\cos 2x+\frac{1}{2}\sin 2|x|\right),\;\; x\in[-\pi/2,\pi/2].
    %\end{align}
Since we are interested in the c.d.f.s, 
    we can use integration by parts followed by algebraic manipulations to obtain (\ref{thm2first}).

    Let us now focus on proving (\ref{thm2second}). To this end, noting that 
    %we observe that
    the distribution of $|\mathbf{v}_j^\dagger \boldsymbol{\Phi}\mathbf{w}_i|^2$ is independent of the indices $i,j$ \cite{Spengler2012}, 
    %Therefore, without loss of generality
    we focus on the case $i=j=1$.  We compute the Laplace transform to obtain the moment generating function (m.g.f.) as
    %\begin{align}
        \[\mathcal{M}(s)=
        %\mathbb{E}\left\{\exp\left(-s\left|\mathbf{v}_1^\dagger \boldsymbol{\Phi}\mathbf{w}_1\right|^2\right)\right\}=
        \mathbb{E}\left\{\exp\left(-\text{Tr}\left[\boldsymbol{\Theta}\mathbf{V}^\dagger \boldsymbol{\Phi}\mathbf{w}_1 \mathbf{w}_1^\dagger \boldsymbol{\Phi}^\dagger \mathbf{V}\right] \right)\right\}\]
   % \end{align}
    where $\boldsymbol{\Theta}=\text{diag}\left(s,0\right)$, and  the expected value is taken with respect to the product measure ${\rm d}\mathbf{V} {\rm d}\mathbf{W}$ where ${\rm d}\mathbf{V}$ and ${\rm d}\mathbf{W}$ are independent and Haar distributed. Therefore,
    by exploiting the independence between $\mathbf{V}$ and $\mathbf{W}$, we  rewrite the m.g.f. as
    \[\mathcal{M}(s)=\mathbb{E}_{\mathbf{W}}\left\{\int_{\mathcal{U}(2)}\exp\left(-\text{Tr}\left[\boldsymbol{\Theta}\mathbf{V}^\dagger \boldsymbol{\Phi}\mathbf{w}_1 \mathbf{w}_1^\dagger \boldsymbol{\Phi}^\dagger \mathbf{V}\right] \right){\rm d} \mathbf{V}\right\},\]
    %\begin{align*}
       % \mathcal{M}(s)=\mathbb{E}_{\mathbf{W}}\left\{\int_{\mathcal{U}(2)}\exp\left(-\text{Tr}\left[\boldsymbol{\Theta}\mathbf{V}^\dagger \boldsymbol{\Phi}\mathbf{w}_1 \mathbf{w}_1^\dagger \boldsymbol{\Phi}^\dagger \mathbf{V}\right] \right){\rm d} \mathbf{V}\right\}
   % \end{align*}
    where $\mathbb{E}_\mathbf{W} (\cdot)$  stands for  the expected value with respect to the Haar measure ${\rm d}\mathbf{W}$.
 Then, we  use \cite[Eq.~89]{James1964ams} to  simplify the inner matrix integral to obtain
 \begin{align*}
 \mathcal{M}(s)=\mathbb{E}_{\mathbf{W}}
       \left\{{}_0\mathcal{F}_0\left(-\boldsymbol{\Theta},\boldsymbol{\Phi}\mathbf{w}_1 \mathbf{w}_1^\dagger \boldsymbol{\Phi}^\dagger\right)\right\}
       \end{align*}
       where
       ${}_0\mathcal{F}_0(\cdot,\cdot)$ is the hypergeometric function of two matrix arguments.
  % \begin{align}
   %\label{eq hypo def}
     %  \mathcal{M}(s)=\mathbb{E}_{\mathbf{W}}
      % \left\{{}_0\mathcal{F}_0\left(-\boldsymbol{\Theta},\boldsymbol{\Phi}\mathbf{w}_1 \mathbf{w}_1^\dagger \boldsymbol{\Phi}^\dagger\right)\right\}
   %\end{align}
    %where ${}_0\mathcal{F}_0(\cdot,\cdot)$ is the hypergeometric function of two matrix arguments. %To facilitate further analysis, 
    We  expand the hypergeometric function as an infinite series to yield \cite{James1964ams}
    \begin{align*}
         \mathcal{M}(s)=\mathbb{E}_\mathbf{W}
         \left\{\sum_{k=0}^\infty \frac{(-1)^k}{k!} \sum_{\kappa} \frac{C_\kappa(\boldsymbol{\Theta})
         C_\kappa(\boldsymbol{\Phi}\mathbf{w}_1 \mathbf{w}_1^\dagger \boldsymbol{\Phi}^\dagger)}{ C_\kappa(\mathbf{I}_2)}\right\}
    \end{align*}
    where $C_\kappa (\cdot)$ denotes the zonal polynomial corresponding to the partition $\kappa=(\kappa_1,\kappa_2)$ with $\kappa_1\geq \kappa_2\geq 0$ and $\kappa_1+\kappa_2=k$.
    Since the  matrices $\boldsymbol{\Theta}$ and $\boldsymbol{\Phi}\mathbf{w}_1 \mathbf{w}_1^\dagger \boldsymbol{\Phi}^\dagger$ have rank one, applying the complex analogue of \cite[ Corollary 7.2.4]{muirhead2005aspects},  we obtain $C_\kappa(\boldsymbol{\Theta})=\text{Tr}^k(\boldsymbol{\Theta})=s$ and $C_\kappa\left(\boldsymbol{\Phi}\mathbf{w}_1 \mathbf{w}_1^\dagger \boldsymbol{\Phi}^\dagger\right)=\text{Tr}^k\left(\boldsymbol{\Phi}\mathbf{w}_1 \mathbf{w}_1^\dagger \boldsymbol{\Phi}^\dagger\right)=1$. Therefore, we obtain 
    \begin{align*}
       \mathcal{M}(s)=\mathbb{E}_\mathbf{W}
         \left\{\sum_{k=0}^\infty \frac{(-s)^k (1)_k}{k!(2)_k}\right\}={}_1F_1(1;2;-s)
    \end{align*}
    where ${}_1F_1 (\cdot;\cdot;\cdot)$ denotes the confluent hypergeometric function of the first kind \cite{gradshteyn2007book} and we have used the fact that $C_k\left(\mathbf{I}_2\right)=(2)_k/(1)_k$ with $(a)_k=a(a+1)\cdots(a+k-1)$ denoting the Pochhammer symbol.
    Finally, using \cite[Eq.~9.211.2]{gradshteyn2007book}, we obtain 
    %Finally, using the integral form of the confluent hypergeomeric function of the first kind \cite[Eq.~9.211.2]{gradshteyn2007book}, we obtain 
    %$\mathcal{M}(s)=\int_0^1 \exp(-st) {\rm d}t$,
     \begin{align}
         \mathcal{M}(s)=\int_0^1 \exp(-st) {\rm d}t
     \end{align}
   from which it can be concluded that the random variable $|\mathbf{v}_1^\dagger \boldsymbol{\Phi}\mathbf{w}_1|^2$ is uniformly distributed over $[0,1]$.
%\begin{remark}
   % When $\mathbf{V},\mathbf{W}\in \mathcal{U}_n$, the above technique can be used to show that $|\mathbf{v}_1^\dagger \boldsymbol{\Phi}\mathbf{w}_1|^2$ is Beta distributed \footnote{A random variable is said to have a beta density function, denoted by $\text{Beta}(a,b)$, if its probability density function is given by
%\begin{align*}
    %f(x)=\frac{\Gamma(a)\Gamma(b)}{\Gamma(a+b)} x^{a-1}(1-x)^{b-1}, \;\;\; 0<x<1,
%\end{align*} 
%where $a,b>0$.} with the parameters $(1,n-1)$.
%\end{remark}
    
   % [Lo] A. Lozano, A. M. Tulino, and S. Verdu ́, “High-SNR power offset in multiantenna communication,” IEEE Trans. Inform.
%Theory, vol. 51, no. 12, pp. 4134–4151, Dec. 2005.
    
    %[Ja] A. T. James, “Distributions of matrix variates and latent roots derived 1307 from normal samples,” Ann. Math. Statist., vol. 35, no. 2, pp. 475–501, 1308 Jun. 1964.
    
    %[Gr] Gradshteyn and Ryzik
    
    % [Mu] R. J. Muirhead, Aspects of Multivariate Analysis.
    
    % [Cs] Composite parameterization and Haar measure for all
%unitary and special unitary groups
%Christoph Spengler,a) Marcus Huber,b) and Beatrix C. Hiesmayr

% [Js] On the singular values of Gaussian random matrices
%Jianhong Shen

%[Er] Erdelyi, Higher Transcendental functions vol. 1.

%\cite{Lozano2005it,James1964ams,muirhead2005aspects,Spengler2012,Shen2001,Erdelyi1953}

\end{appendices}
%\vspace{-1mm}
\bibliographystyle{IEEEtran}
\bibliography{reference,IEEEabrv}

\end{document}